\documentclass[12pt,preprint]{aastex}
\usepackage{graphicx}
\usepackage{epstopdf}
\usepackage{xcolor}
 \usepackage{epsfig}

\newcommand{\eb}{\begin{equation}}
\newcommand{\ee}{\end{equation}}
\newcommand{\be}{\begin{equation}}
\newcommand{\ba}{\begin{eqnarray}}
\newcommand{\ea}{\end{eqnarray}}

\shorttitle{Tidal evolution of TRAPPIST-1 planets}
\shortauthors{Makarov et al.}
\begin{document}

\title{\large{Spin-orbital tidal dynamics and tidal heating\\ in the TRAPPIST-1 multi-planet system}\vspace{3mm}}
\author{
                         {\large{Valeri V. Makarov}}\\
                         {\small{US Naval Observatory, Washington DC 20392}}\\
                         {\small{e-mail: ~valeri.makarov$\,$@$\,$navy.mil~$\,$}}\vspace{2mm}\\
 ~\\
                         {\large{Ciprian T. Berghea}}\\
                         {\small{US Naval Observatory, Washington DC 20392}}\\
                         {\small{e-mail: ~ciprian.berghea$\,$@$\,$navy.mil~$\,$}}\vspace{2mm}\\
 ~\\
                         {\large{Michael Efroimsky}}\\
                         {\small{US Naval Observatory, Washington DC 20392}}\\
                         {\small{e-mail: ~michael.efroimsky$\,$@$\,$navy.mil~$\,$}}\vspace{2mm}
}

% \affil{US Naval Observatory, Washington DC 20392}
% \email{valeri.makarov@navy.mil~,~~~ciprian.berghea@navy.mil~,~~~michael.efroimsky@navy.mil}

\begin{abstract}
We perform numerical simulations of the TRAPPIST-1 system of seven exoplanets orbiting a nearby M dwarf, starting with a previously suggested stable configuration. The long-term stability of this configuration is confirmed, but the motion of planets is found to be chaotic. The eccentricity values are found to vary within finite ranges. The rates of tidal dissipation and tidal evolution of orbits are estimated, assuming an Earth-like rheology for the planets. We find that under this assumption, the planets $\,b\,$, $\,d\,$, $\,e\,$ were captured in the 3:2 or higher spin-orbit resonances during the initial spin-down, but slipped further down into the 1:1 resonance. Depending on its rheology, the innermost planet $\,b\,$ may be captured in a stable pseudosynchronous rotation. 
Nonsynchronous rotation ensures higher levels of tidal dissipation and internal heating. The positive feedback between the viscosity and the dissipation rate~---~and the ensuing runaway heating~---~are terminated by a few self-regulation processes. When the temperature is high and the viscosity is low enough, the planet spontaneously leaves the 3:2 resonance. Further heating is stopped either by passing the peak dissipation or by the emergence of partial melt in the mantle. In the post-solidus state, the tidal dissipation is limited to the levels supported by the heat transfer efficiency. The tides on the host star are unlikely to have had a significant dynamical impact. The tides on the synchronized inner planets tend to reduce these planets' orbital eccentricity, possibly contributing thereby to the system's stability.
\end{abstract}


 \section{Introduction  \label{intro.sec}}
 \label{firstpage}

 Main-sequence M dwarfs are a numerically dominant stellar species in the Galaxy, and the possibility of finding a terrestrial planet in their narrow habitable zones provides the best chance of locating a world suitable for biological life in the
relative vicinity of the Sun. Super-Earth exoplanets in the habitable zones of nearby M dwarfs are probably
abundant, with an estimated rate of $\eta_{\rm Earth}=0.41^{+0.54}_{-0.13}$ per host
star \citep{bonf}. Super-Earths are
more massive than the Earth, but smaller than 10 Earth masses. The recent discovery of the TRAPPIST-1 system of seven planets \citep{gil}
transiting a nearby, otherwise unremarkable M dwarf \citep{rei,fil,gil16} opens a new realm of extrasolar worlds --
tightly packed systems of Earth-sized planets. The estimated masses of planets in this system range from $0.4$
to $1.4\,M_{\rm Earth}$. With the currently outermost planet orbiting the star in less than 19 days, we are challenged to explain
how such compact systems emerge and why they remain stable on the astronomical time scale.

 A planet's rotation is one of the determinants for the presence and properties of its atmosphere, and for the distribution of the stellar irradiation on the surface. The climate pattern may be radically different on planets in the habitable zone, depending on the spin-orbit resonance they reside in \citep{wan}. Physically justified, frequency-dependent tidal models, such as those based on the Maxwell or combined Maxwell--Andrade rheologies \citep{efr1,efr2}, predict high probabilities of capture into higher-than-synchronous spin-orbit resonances for close-in exoplanets with finite eccentricities \citep{ma12}. This result holds for a wider range of realistic models, such as the Newtonian creep model developed by \citet{fer}. The orbital eccentricity and average viscosity are the two crucial parameters defining these probabilities.

 The issue of dynamical stability and long-term survival of a tightly packed close-in system such as TRAPPIST-1 has already been discussed in the discovery paper by \citet{gil}. A system with the observed set of parameters breaks up within a relatively short time in numerical $N$-body integrations. Fine-tuning of initial orbital parameters is
 required to find a long-term stable configuration within the observational uncertainties. This indicates that the multi-dimensional parameter space is shredded, with small islands of stable solutions separated by areas of instability. How can a real physical system of high complexity converge on a small island of stability among an infinite number of possible unstable configurations? The stability islands should be attractors, i.e., there should exist a regulatory force driving the system to one of such islands and damping small perturbations. \citet{gil} suggested that considerable tidal effects can provide the required stabilization. However, tidal forces act much too slowly even for such close planets, while the break-up of an unstable system happens well within 1 Myr. Even though tidal effects are important on the lifetime scale, we still have to explain how the system originally found a stable configuration. \citet{tam} proposed that a convergent migration in a protoplanetary disk can be such a regulatory mechanism. Subject to dynamic friction in the disk, young planets migrate generally inward and toward smaller eccentricities. Still, something would halt this process when a planet reaches a stability niche, even before the disk dispersal. This may be possible when the stability niche is also associated with a certain mean-motion resonance (MMR), or better still, a system of MMRs with other planets, because an MMR is known to be an attractor. The surviving protoplanets end up in intricate patterns of near-commensurability \citep{ter}, a class of $N$-body systems radically different from the Solar System. \citet{lug} find that the TRAPPIST-1 system represents a chain of MMRs, where each triplet of adjacent
 planets is locked in a three-body resonance.

 In this paper, we begin with testing a long-term stable configuration suggested by \citet{qua}. We also test the system for chaos, using the siblings method in numerical integrations. The resulting orbital eccentricities for the seven planets, termed for brevity as TR1-1, TR1-2, \ldots, TR1-7 (instead of the traditional notation TRAPPIST-1 $b$, $c$, \ldots, $h$), are used to estimate the likely spin-orbit end-states of the planets, assuming an Earth-like rheology described by the Maxwell model. We show that not all of the planets are expected to fall in the 1:1 resonance (synchronous rotation) if they remain cold and inviscid during the initial spin-down.
 The innermost planets, in particular, are more likely to be initially locked in higher resonances (2:1 or 3:2) despite their low mean eccentricity.
 This gives a significant boost to the estimated rate of energy dissipation. The rate of tidal heating is so high for TR1-1 that one may wonder if the planet should be completely molten. We further describe the mechanism of tidal heating self-regulation, which could prevent a planet with a rocky mantle from going past solidus. Besides defining the tidal heating rates, the rotation states of tidally deformed bodies influence their orbital evolution. It should not be taken for granted that the tidal forces always circularize and shrink orbits. We show that the tides in a fast-rotating host star may have the opposite effect on the outer planets' orbits.

 Our research is based on analytical and semianalytical techniques. Applicable only to uniform or two-layered bodies, the analytical and semianalytical techniques do not lead to a self-consistent detailed modeling of internal processes, such as the internal heat transfer or the surface irradiation.
 In contradistinction, numerical simulations of tidal dissipation for specific template cases provides certain advantages. They allow for a more detailed investigation into the effects of a multilayered structure, and make it easier to explore the influence of particular rheologies, such as Maxwell \citep{bou,wal}, Burgers \citep{hen}, and Kelvin--Voigt \citep{zle,fro}. However, these advantages of numerics are all but gone in view of the great uncertainties we meet when analyzing the tidal scenarios for exoplanets. We are limited to an approximate, order-of-magnitude estimation based on a wide range of possible values for the most significant parameters, not to mention the remaining ambiguity of physical modeling of the layers. For this reason, useful insight can be gained through a low-cost analytical or semi-analytical approach; and hypotheses instilled in this approach can be tested against the observational data.

 \section{The orbits of the TR1 planets \label{orb.sec}}

 For our analysis of the orbital stability of the TR1 system, we used observational data sets from \citet{gil} and the configuration reported by
 \citet[][Table 3]{qua} as long-term stable. The integrations, which used the observed orbital parameters as initial conditions, invariably broke apart after $\,10^5$--$10^6$ years. However, the configuration presented by \citet{qua} remained intact for the duration of integration (which was 10 Myr in our case). Among the initial configurations tested by those authors, 28.2\% proved to be stable. We confirm that relatively small changes in the initial values of the orbital elements can result in a stable system, suggesting the small size of the stability islands in the sea of unstable configurations.

\begin{figure}[htbp]
  \centering
  \includegraphics[angle=0,width=0.95\textwidth]{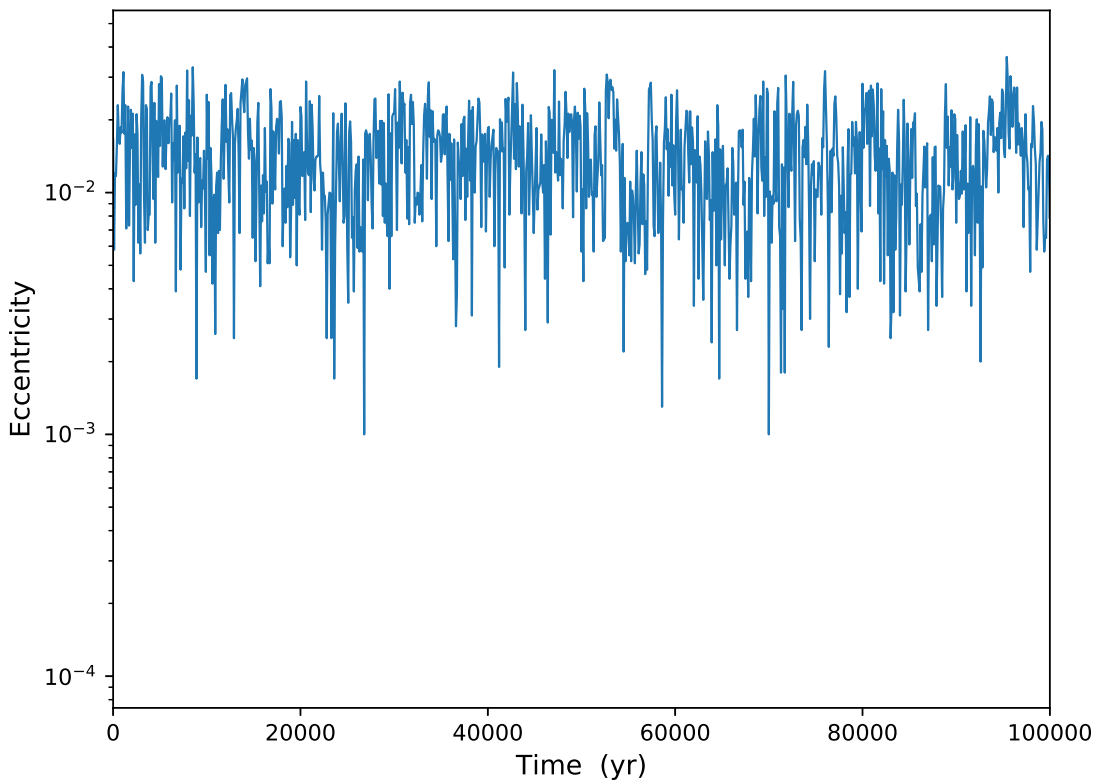}
\caption{Numerically integrated behavior of planet h (TR1-7) eccentricity. \label{e.fig}}
\end{figure}

 We find that the eccentricity of each planet rapidly varies in an apparently stochastic way, but remains within a certain range
 for the entire duration. Figure ~\ref{e.fig} shows the result for the outermost planet TR1-7. Other derived characteristics are given
 in Table \ref{e.tab}, including the mean, median, and robust standard deviation. The latter values, for example, are found in the relatively
 narrow range $[0.004,\;0.009]$, and are strongly correlated with the mean (or robust mean) values. The mean eccentricities are
 all rather small, with the largest value reaching 0.017 for planet TR1-3, but they do differ by more than a factor of 2
 among the planets. Such differences are quite significant for the tidal evolution and equilibrium states of the planets.
\begin{deluxetable}{rrrrrrr}
\tablecaption{Numerically simulated statistics for orbital eccentricity of TRAPPIST-1 planets.
\label{e.tab}}
\tablewidth{0pt}
\tablehead{
\multicolumn{1}{c}{Planet}  & \multicolumn{1}{c}{Range} & \multicolumn{1}{c}{Mean} & \multicolumn{1}{c}{Median} & \multicolumn{1}{c}{StD} & \multicolumn{1}{c}{Robust Mean} & \multicolumn{1}{c}{Robust StD}}
\startdata
1\dotfill $b$ & $0.0365$ & $0.0112$ & $0.0107$ & $0.00564$ & $0.0113$ & $0.00585$\\
2\dotfill $c$ & $0.0289$ & $0.00874$ & $0.00830$ & $0.00441$ & $0.00875$ & $0.00455$\\
3\dotfill $d$ & $0.0587$ & $0.0171$ & $0.0163$ & $0.00863$ & $0.0172$ & $0.00895$\\
4\dotfill $e$ & $0.0454$ & $0.0138$ & $0.0131$ & $0.00694$ & $0.0138$ & $0.00720$\\
5\dotfill $f$ & $0.0321$ & $0.0102$ & $0.00970$ & $0.00508$ & $0.0101$ & $0.00525$\\
6\dotfill $g$ & $0.0238$ & $0.00749$ & $0.00710$ & $0.00376$ & $0.00750$ & $0.00390$\\
7\dotfill $h$ & $0.0419$ & $0.0121$ & $0.0115$ & $0.00611$ & $0.0121$ & $0.00630$\\
\enddata
 % \tablecomments{}
\end{deluxetable}

Due to lack of better estimates, we started numerical integrations with all inclinations set to zero. This does not seem to
 have constrained the results because the inclinations, like the eccentricities, undergo rapid and apparently stochastic
 changes within small ranges. This may be the manifestation of small but persistent random-walk-like excitation of orbital
 tilts by the neighbor planets. A finite inclination gives rise to latitude tides and additional dissipation. The tidal interaction
 also changes the equator's obliquity through the conservation of angular momentum. The combination of externally excited inclination
 and tidally driven precession of the equator can considerably change the spectrum of possible spin-orbit resonances \citep{bou}.
 In this paper, this complication is ignored and the consideration is limited to the 1D case by nominally assuming zero values
 for the inclinations.

 The semimajor axes display some jitter too, but the variations are small and of no significance for the tidal considerations.
 Their mean values are also very close to the initial values in Table 3 of \citet{qua}. Overall, this system proves to be
 long-term stable despite jitter variations of orbital elements, because the eccentricity is staying within a small range constrained by
 some regulatory factor in its dynamics. Remarkably, as we noted before, a system with the nominal observed parameters
 is not stable at all. This is similar to the state of the Solar System, where long-term stable and unstable configurations
 are tightly spaced within the observational uncertainties of the planetary ephemerides \citep{hay1, hay2}. This is
 explained by the intrinsic chaos in the motion of solar-system bodies.

 \begin{figure}[htbp]
 \centering
   \includegraphics[height=11cm]{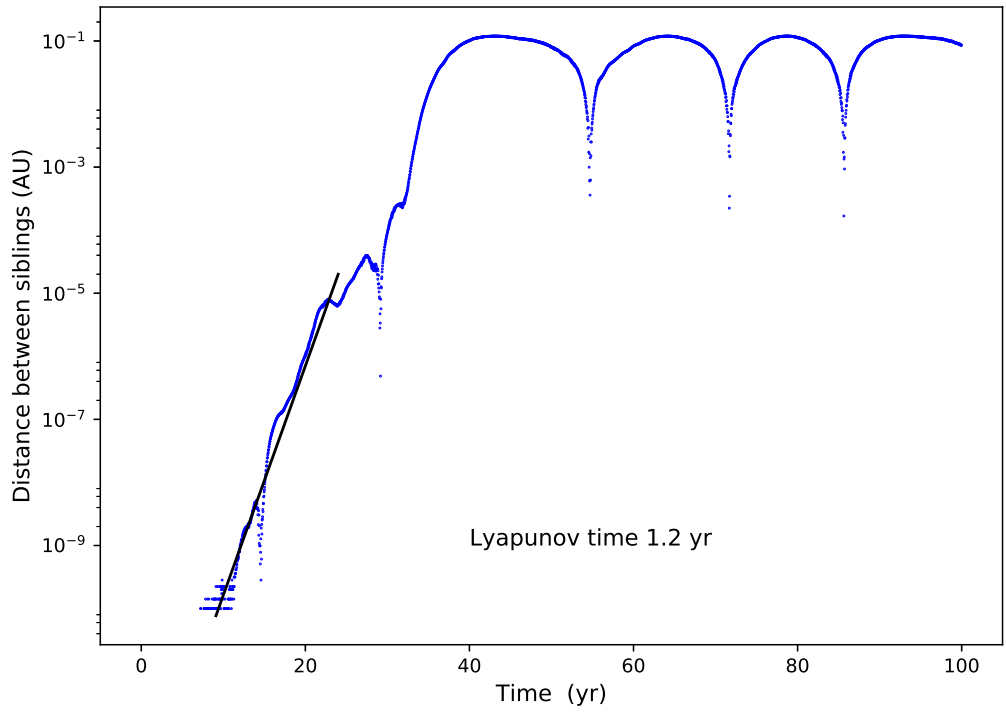}
 % \includegraphics[angle=0,width=0.89\textwidth]{siblings.eps}
 % \plotone{siblings.eps}
  \vspace{3.4cm}
 \epsscale{0.75}
 \caption{Simulated distance in semimajor axis $a$ between two sibling trajectories for planet TR1-7.}
 \label{sib.fig}
 \end{figure}

 Using the
 % symplectic integrator HNBody, version 1.0.7, \citep{rauch} with the hybrid symplectic option, and
 the symplectic Bulirsch--Stoer option of the Mercury code \citep{cha} and a time step of 0.025 d, we integrated two sibling trajectories whose initial semimajor axis values had a relative difference of  $10^{-14}$. The absolute distance in astronomical units between the siblings was then computed as a function of time. Chaotic motion manifests itself as an exponential divergence between the trajectories. We find such exponential divergence in the trajectories of all seven planets. Figure \ref{sib.fig} demonstrates this result for planet TR1-7. The rate of divergence allows us to estimate the Lyapunov time, which is approximately 1.2 yr for this planet. The conclusion is that even a tiny change or uncertainty in the orbital parameters makes the long-term prediction of the future state impossible.

 In our previous simulations of exoplanet systems, we found clear signs of chaos for GJ 581 \citep{mbe}, but not for GJ 667C \citep{mabe}. The former is a complex system with a few planets interacting, whereas the latter is assumed to include only two planets locked in the 2:1 MMR. Chaotic behavior may be common for systems with more than two planets, even when some of them are locked in mutual resonances.

\begin{deluxetable}{lrr}
\tablecaption{Maximum tidal heating rates and the peak Maxwell times.\label{peak.tab}}
\tablewidth{0pt}
\tablehead{
\multicolumn{1}{c}{Planet}  &
\multicolumn{1}{c}{$dE/dt$} & \multicolumn{1}{c}{$\tau_M$}\\
\multicolumn{1}{c}{}  & \multicolumn{1}{c}{W} & \multicolumn{1}{c}{d}\\}
\startdata
TR1-1 & \dotfill $3.7\times 10^{20}$ & $0.048$\\[1ex]
TR1-2 & \dotfill $4.9\times 10^{16}$ & $0.14$\\[1ex]
TR1-3 & \dotfill $3.5\times 10^{17}$ & $0.035$\\[1ex]
TR1-4 & \dotfill $1.9\times 10^{17}$ & $0.086$\\[1ex]
TR1-5 & \dotfill $9.3\times 10^{12}$ & $0.19$\\[1ex]
TR1-6 & \dotfill $6.0\times 10^{14}$ & $0.76$\\[1ex]
TR1-7 & \dotfill $5.3\times 10^{13}$ & $0.20$\\[1ex]
\enddata
\end{deluxetable}

 \section{The Maxwell model and the tidal quality function $\,k_2/Q$ \label{kva.sec}}

 In the Solar System, Mercury is the closest analogue to the TR1 planets, in the sense of orbital separation and mass. Mercury is believed to have a massive molten core, whose existence is revealed by the large amplitude of forced libration detected via ground-based radar measurements \citep{mar}. The core should be nearly decoupled from the solid mantle (i.e., it should have a low coefficient of friction) to enable the increased libration. Unlike the TR1 planets, however, Mercury's orbital eccentricity is quite high, which explains its permanent 3:2 spin-orbit resonance. Early attempts to explain this state by using simplistic tidal models (such as the constant-$Q\,$ or constant-time-lag models) met considerable difficulties, rendering low values for the probability of any equilibrium state apart from the synchronous rotation, the 1:1 spin-orbit resonance \citep{corla04}. \citet{noel} demonstrated that these difficulties are resolved within a physics-based -- and, therefore, more realistic -- description of the rheological response.
 Based on a combined Maxwell--Andrade rheology introduced on physical grounds by \citet{efr1,efr2}, the tidal response demonstrated a pronounced frequency dependence. Armed with this model, and employing physically reasonable ranges of the parameters' values, \citet{noel} demonstrated that the 3:2 resonance is indeed the most likely outcome of the tidal evolution of a homogeneous rotator, and that the capture is a very quick a process (less than 20 Myr). On the other hand, the presence of a massive decoupled core made the higher resonances, such as 5:2 or 2:1, more probable end states. These results led the authors to the conclusion that, in all likelihood, Mercury was captured in the 3:2 spin-orbit state shortly after its accretion, i.e., well before it was heated up by radioactivity. The authors also acknowledged the possibility that Mercury might have been trapped in one of such high resonances, before a massive impact during a period of relatively low eccentricity drove it out of that state.

\begin{figure}[htbp]
 \plotone{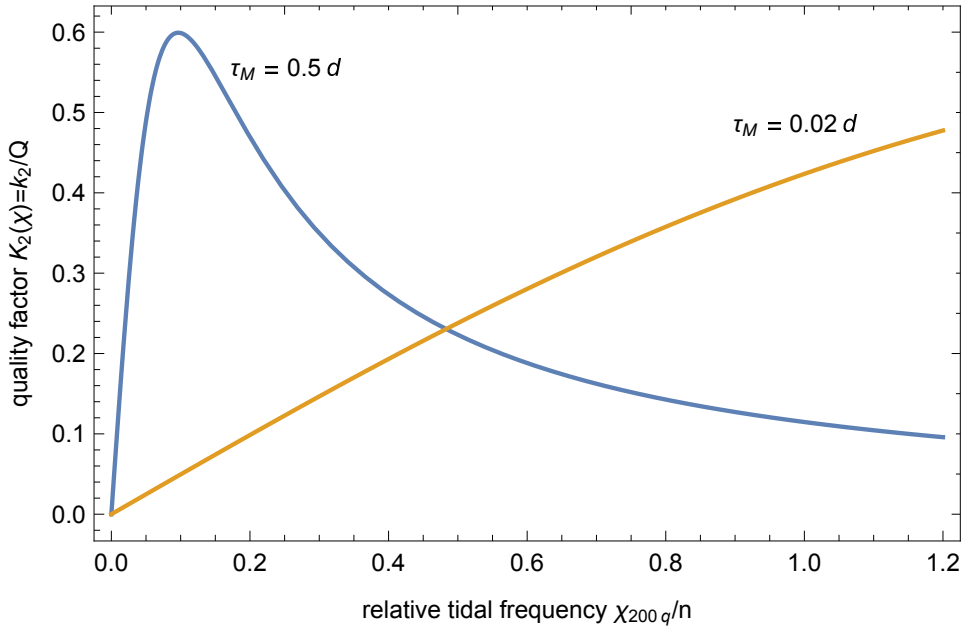} \epsscale{0.75}
 \caption{Tidal quality ratio $\,K=k_2/Q\,$ for the planet TR1-1, as a function of the tidal excitation frequency, plotted for two
 values of the Maxwell time, $\tau_M=0.02$  and $0.5$ d.
 \label{kva.fig}}
 \end{figure}

 The Maxwell model has long been in use in the studies of exoplanets and the solar-system bodies \citep[e.g.,][]{sto,cor}. The parameters entering this model~---~rigidity and viscosity~---~cannot be directly constrained by observation. It is also difficult to estimate them theoretically, because these parameters are sensitive to the presence of partial melt, to the pressure, and even more so to the temperature.
 The high temperatures and pressures required to emulate the conditions of a deep mantle are hard to achieve in laboratory measurements, and tidal manifestations remain the principal method for the indirect estimation of these parameters. Furthermore, the tidal response of terrestrial planets, satellites, and small bodies turns out to be less sensitive to the rigidity and much more sensitive to the viscosity \citep{efr3}. Figure \ref{kva.fig} shows the dimensionless tidal quality function, $\,k_2/Q\,$, plotted against the excitation frequency for two values of the Maxwell time of the TR1-1 planet.\,\footnote{~The Maxwell time is the ratio of the mean shear viscosity of the mantle to its mean shear rigidity modulus, $\,\tau_M=\eta/\mu\,$.\label{1}} For the shorter Maxwell time (implying a lower viscosity value), the dependence is almost linear, and the classic constant-time-lag model turns out to be a good approximation. For somewhat colder mantles of higher viscosity, the function is remarkably nonlinear,
 and conclusions can be drawn only from numerical simulations.

 \section{Self-regulation of tidal heating}

 As the partners tidally interact with one another over billions of years, tides become an important factor of the dynamical history of the system. The rate of tidal dissipation depends on a number of orbital and planetary parameters, including the eccentricity $e$, the masses of the star and the planet ($M_1$ and $M_2$), the semimajor axis $a$, the planet's radius $R$, and its Maxwell time $\tau_M$. The dependences of the tidal evolution rate and of the capture probabilities on these parameters are quite strong and generally nonlinear, making the physical interpretation of observations intrinsically uncertain. The approach we are using here is to assume some of the parameters to be close to the Earth's benchmark values when available, and to consider a wide range of values otherwise.

\begin{figure}[htbp]
\plotone{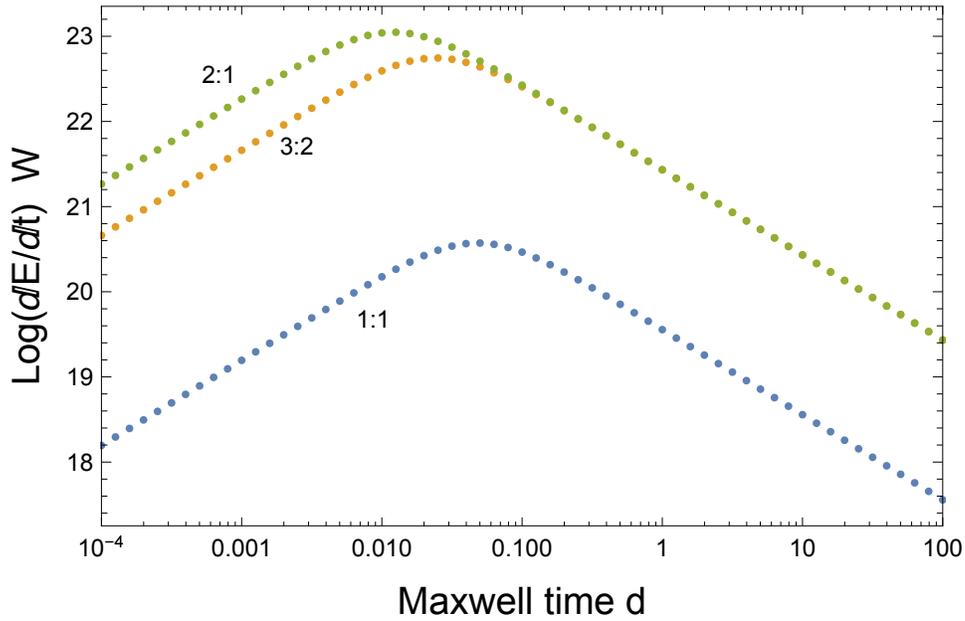} \epsscale{0.75}
\caption{Estimated rate of tidal energy dissipation for planet TR1-1 as a function of the Maxwell time $\tau_M$, for three
possible spin-orbit resonances, 1:1, 3:2, and 2:1.   \label{dedt.fig}}
\end{figure}

 Figure \ref{dedt.fig} displays the Maxwell-time dependence of the energy dissipation rate (i.e., of the power exerted by the tidal friction) in the planet TR1-1, for a fixed orbital period $P_{\rm orb}=1.51$ d. We used the formula for the tidal dissipation rate derived by \citet{efma}, which is a generalization of a result obtained earlier by \citet{peca}. While \citet{peca} were dealing exclusively with a synchronous case, \citet{efma} extended their theory to an arbitrary spin state, resonant or nonresonant.
 Due to this planet's considerable size and its proximity to the star, the tidal dissipation in this planet is very intense. It surpasses by many orders of magnitude the tidal dissipation rates typical for the solar-system bodies. The peak rate for the 1:1 spin-orbit resonance, which is the likeliest spin state (see Section \ref{c.sec}), reaches
 $4\times 10^{20}$ W. This results in an equilibrium surface flux of $6.6\times 10^5$ W m$^{-2}$. By comparison, the thermal flux on Earth, caused by the tides from the Moon and the Sun, amounts to only $\simeq 0.001$ W m$^{-2}$ \citep{henh}. As can be seen from the plot, the Maxwell time corresponding to the maximal power is 0.048 d. The maximal dissipation rates and the corresponding values of $\tau_M$ are given in Table \ref{peak.tab} for the seven planets, under the synchronous rotation assumption.

 \section{The likely spin--orbit states of the TR1 planets \label{c.sec}}

 The rotation history of a tidally perturbed elongated body is governed by the torque $\,\vec{\mathcal{T}}^{^{(TRI)}}$ due to the body's permanent triaxiality and the torque $\,\vec{\mathcal{T}}^{^{(TIDE)}}\,$ due to tides in the body \citep{danb,gold,gol68}.  At small obliquities (i.e., when the perturber's orbital plane is close to the equatorial plane of the tidally perturbed body), only the polar components of the torques matter, and the equation of motion acquires the form
 \ba
 \stackrel{\bf\centerdot\,\centerdot}{\theta~}\,=~\frac{\,{\cal{T}}^{\rm{^{\,(TRI)}}}_{polar}\,+~{\cal{T}}^{\rm{^{\,(TIDE)}}}_{polar}}{C~~}~\,
 =~\frac{{\cal{T}}^{\rm{^{\,(TRI)}}}_{polar}\,+~
 {\cal{T}}^{\rm{^{\,(TIDE)}}}_{polar}}{\xi ~M~R^{\,2}}\,~,
 \label{eq.eq}
 \label{201}
 \label{eq:despinning}
 \ea
 Here $\,M\,$, $\,R\,$ and $\,C\,=\,\xi\,M\,R^2\,$ are the mass, radius and the maximal moment of inertia of the body, while $\,\xi\,$ is a dimensionless factor (equal to $\,2/5\,$ for a homogeneous sphere). The rotation angle $\,\theta\,$ of the body is conventionally reckoned from the line of apsides to the largest elongation axis.
 
 The permanent-triaxiality-generated torque $\,\vec{\mathcal{T}}^{^{(TRI)}}$ is oscillating, while the tidal torque $\,\vec{\mathcal{T}}^{^{(TIDE)}}\,$ contains both a secular and an oscillating parts \citep{efr1}. The long-term evolution and the likely end states of the rotator are defined mainly by the interplay of $\,\vec{\mathcal{T}}^{^{(TRI)}}$ and the secular part\footnote{~While the oscillating part of the tidal torque changes the fate of each particular trajectory, the overall statistics of tidal capture in spin-orbit resonances is defined overwhelmingly by the secular part \citep{mbe}.} of $\,\vec{\mathcal{T}}^{^{(TIDE)}}\,$.

 It is commonly accepted that a planet rapidly rotates at the time of formation, whether in the prograde or retrograde sense. The gradual dissipation of the rotational energy by the tidal friction causes angular deceleration. For close-in planets, the energy damping rate is intensive enough to make the spin-down time short (e.g., $10 - 20$ Myr for Mercury). Still, it is not sufficiently intensive to cause a significant tidal heating, because the overall supply of the rotational energy is limited. Simple calculations show, for example, that the lengthening of the solar day on the Earth results in a heating much smaller than the surface heating by the solar irradiation or than the internal heating by the radioactive decay. So the planet quickly reaches a spin-orbit equilibrium state before having undergone a significant internal restructuring. The subsequent history of the planet after the capture much depends on which equilibrium spin-orbit state it has ended up in.

\begin{figure}[htbp]
\plotone{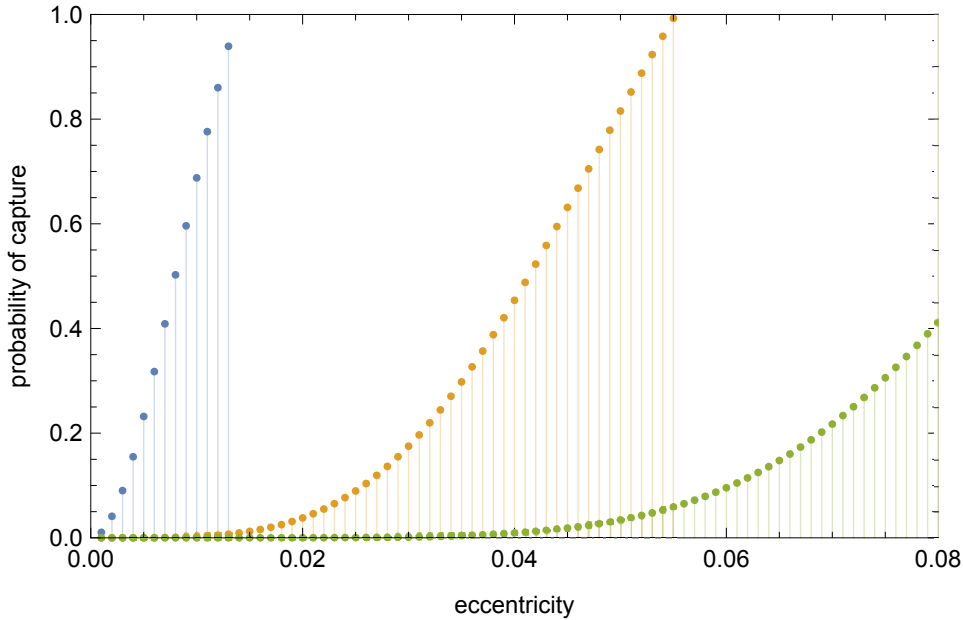} \epsscale{0.75}
\caption{Probabilities of capture of planet TR1-1 into spin-orbit resonances 3:2, 2:1, and 5:2 (left to right)   \label{prob.fig}}
\end{figure}

 It has often been assumed in the literature, incorrectly, that all close-in exoplanets are tidally synchronized. Facing the stars with one side for billions of years, such planets should be quite limited in their ability to support a highly developed biological life, or even a stable atmosphere rich in volatiles. This paradigm has to be reconsidered, however. With the example of two nearby systems, GJ 581 and 667, harboring potentially habitable super-Earths, \citet{mbe,mabe} posited that the eccentricities can be large enough for these planets to be captured in a 3:2 or even higher order spin-orbit resonance. This leads to a much more uniform distribution of stellar irradiance on the surface, but also to increased tidal heating. Our task now is to estimate how likely it may be for the TR1 planets to be captured in such Mercury-like resonances.
%\begin{figure}[htbp]
%  \centering
%  \plottwo{25kink.eps}{25capture.eps}
%\caption{Left: Secular tidal acceleration of GJ 667Cc versus normalised spin rate in the vicinity of the 5:2
%spin-orbit resonance. Right: A simulated capture of GJ 667Cc into the 5:2 resonance.\label{kink.fig}}
%\end{figure}

 \begin{deluxetable}{lrr}
 \tablecaption{Probabilities of the TRAPPIST-1 planets' capture in higher-than-synchronous spin--orbit resonances, provided all these planets' Maxwell time and dynamical triaxiality assume the values $\,\tau_M=1$ d \,and $\,(B-A)/C=1.9\times 10^{-5}\,$.\label{prob.tab}}
 \tablewidth{0pt}
 \tablehead{
 \multicolumn{1}{c}{Planet}  & \multicolumn{2}{c}{Probability} \\
 \multicolumn{1}{c}{} & \multicolumn{1}{c}{3:2} &
 \multicolumn{1}{c}{2:1}\\}
 \startdata
TR1-1 & $1$ & $0.195$\\[1ex]
TR1-2 & $0.02$ & $0$\\[1ex]
TR1-3 & $0.681$ & $0.014$\\[1ex]
TR1-4 & $0.693$ & $0.024$\\[1ex]
TR1-5 & $0$ & $0$\\[1ex]
TR1-6 & $0.05$ & $0$\\[1ex]
TR1-7 & $0.017$ & $0$\\[1ex]
\enddata
\end{deluxetable}

 An early formula for the probability of capture in a spin-orbit resonance was proposed by \citet{gol68} for a CPL (constant phase-lag) model, i.e., for a model wherein the intensity of tidal interaction was frequency-independent. The formula was generalized by \citet{ma12} for a general frequency-dependent tidal model. Here we used the latter development to compute capture probabilities for a range of values of $\,\tau_M\,$, the most uncertain parameter. The rheology was presumed to be Maxwell~---~which should be adequate for close-in exoplanets of terrestrial composition.\,\footnote{~For physical reasons, the Maxwell model
 should be substituted with a combined Maxwell-Andrade model, when the forcing frequencies are higher than a certain threshold \citep{efr1}. Exponentially sensitive to the temperature, the value of this threshold is of the order of $\,yr^{-1}\,$ for the Earth, but is much higher for warmer planets \citep[][Eq. 17]{kar0}. So for close-in planets (presumed sufficiently warm) the Andrade mechanism may be neglected and the tidal reaction of the mantle may be presumed Maxwell.\label{3}}

 Figure \ref{prob.fig} shows the results for the innermost planet TR1-1, which is subject to the strongest tidal interaction, versus eccentricity, with a fixed Maxwell time of 1 d. Depending on the value of eccentricity at the time of capture, the planet might have been trapped in any of the higher-than-synchronous resonances, 3:2, or 2:1, or higher. It is, however, doubtful that for a dynamically stable configuration
 the eccentricity could be much larger than its present value (0.011). A tightly packed multiplanet system with initial high eccentricities would be unlikely to survive in its present-day configuration. Therefore, using the fixed values emerging from extended orbital integrations, e.g., the robust means shown in Table \ref{e.tab},
  is justified. Table \ref{prob.tab} renders the thus estimated probabilities of capture into the 3:2 and 2:1 spin-orbit states. We determine that planets TR1-2, TR1-5, TR1-6, and TR1-7 should almost certainly be synchronized. On the other hand, planets TR1-1, TR1-3, and TR1-4 are likely to have been initially captured in the 3:2 resonance, with a finite probability of an even higher spin state.

\begin{deluxetable}{lr}
\tablecaption{Explanation of notations \label{nota.tab}}
\tablewidth{0pt}
\tablehead{
\multicolumn{1}{c}{Notation}  &
\multicolumn{1}{c}{Description}\\
}
\startdata
$R$ & \dotfill the radius of a planet \\
$\theta$ & \dotfill the polar rotation angle of a planet \\
$\dot{\theta}$ & \dotfill the polar rotation rate of a planet \\
$\theta_1$ & \dotfill the polar rotation angle of the star \\
$\dot{\theta}_1$ & \dotfill the polar rotation rate of the star \\
$M_2$ & \dotfill the mass of a planet \\
$M_1$ & \dotfill the mass of the star \\
$a$ & \dotfill the semimajor axis of a planet \\
$r$ & \dotfill the instantaneous distance of a planet from the star \\
$\nu$ & \dotfill the true anomaly of a planet \\
$e$ & \dotfill orbital eccentricity \\
$M$ & \dotfill the mean anomaly of a planet \\
$i$ & \dotfill the inclination of the orbit on a planet's equator \\
$i_1$ & \dotfill the inclination of the orbit on the star's equator \\
$A\leq B < C$ & \dotfill the moments of inertia of a planet\\
$\xi$ & \dotfill the moment of inertia coefficient \\
${\cal{A}}_2$ & \dotfill the dimensionless rigidity of a planet \\
$n$ & \dotfill the mean motion of a planet $\,(n = 2\pi/P_{\rm orb})$ \\
${G}$ & \dotfill the gravitational constant, $=66468$ m$^3$ kg$^{-1}$ yr$^{-2}$ \\
$\eta$ & \dotfill the mean shear viscosity of a planet\\
$\mu$ & \dotfill the mean shear rigidity modulus of a planet\\
$\tau_M\,\equiv\,\eta/\mu$ & \dotfill the mean Maxwell time of a planet\\
$\rho$ & \dotfill the mean density of a planet\\
\enddata
\end{deluxetable}

\begin{figure}[htbp]
  \centering
  \plotone{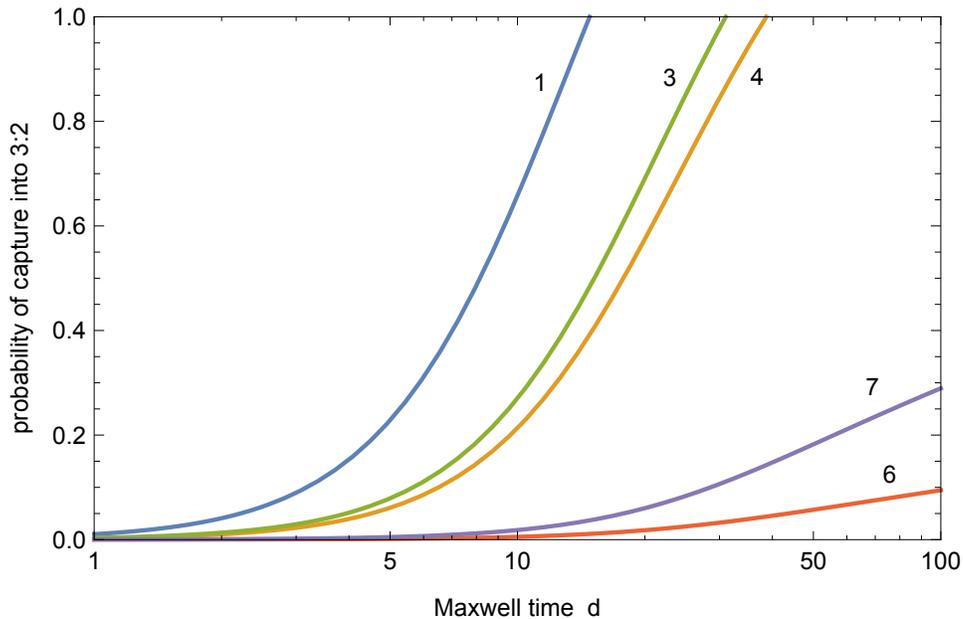}
\caption{Probabilities of capture of the TRAPPIST-1 planets in the 3:2 spin-orbit resonance versus the Maxwell time. \label{pcapt.fig}}
\end{figure}

 The largest uncertainty in this analysis comes from the assumed Maxwell time. This parameter, poorly known even for the solar-system bodies, is a strong function of the mantle's temperature and pressure through its proportionality to viscosity (see Section \ref{sol.sec} below).
 The inner planets of the system receive vast amounts of tidal heating and should be hot, lowering the Maxwell time. Figure \ref{pcapt.fig}
 shows how the probability of capture in the 3:2 spin-orbit resonance depends on the effective Maxwell time. Cold planets have higher viscosity and are more readily captured into this resonance. Planets TR1-1, 3, and 4 are certainly captured if their Maxwell time is longer than $\,\sim 40$ d. The outermost planets 6 and 7 can be captured, in principle, if they were cold (like the Earth today) at the time of their initial spin-down. Planets 2 and 5 cannot be in this resonance because of their low eccentricities.

 Apart from the strong dependence on the Maxwell time $\,\tau_M\,$ and eccentricity $\,e\,$, the probabilities of capture also depend on the triaxiality $\sigma=(B-A)/C$, which is unknown. This dependence, however, is relatively weak. Near-oblate bodies (i.e., those with a vanishingly small $\,(B-A)/C\,$) are always captured into a resonance if in its vicinity the secular part of the tidal torque is positive (accelerating in the prograde sense). A rule to keep in mind is that less triaxial planets are more readily captured into higher-than-synchronous resonances, with all other parameters being equal. Based on the data for the Solar System, larger planets are less triaxial than the smaller planets or moons. The triaxiality value we assumed here, $\,(B-A)/C=1.9\times10^{-5}\,$, is close to that of the Earth \citep{lam} and therefore is a reasonable educated guess.

High-energy impact is a stochastic factor that may also define the spin-orbit states of exoplanets. Within some scenarios, the current 3:2 spin-orbit state of Mercury could have been the outcome of such impacts in the past \citep{cola12,noel}. Speculatively, early collisions at the stage of planetary system formation could not have a long-lasting effect because the eccentricities have settled to finite values, and the tidal mechanisms had sufficient
time to change the spins of the TR1 planets. If a late bombardment episode were to happen in this settled state, fortuitous tangential hits
could have temporarily driven synchronously rotating planets into higher spin-orbit resonances, causing periods of increased internal heating.

The inner TR1 planets may also be subject to considerable retrograde torques originating from their hypothetical atmospheres. Owing to the friction
coupling between the atmosphere and the planet's surface \citep{dobr}, the secular thermal torque on the atmosphere is transferred to the planet,
counterbalancing the action of the lagging solid-body bulge. This is believed to explain the retrograde rotation of Venus with its
massive and thick atmosphere. \citet{lec} discussed that close exoplanets orbiting low-mass stars may also be dominated by thermal atmospheric
tides (even if their atmospheres are thin) and may maintain a stable asynchronous rotation. We defer analysis of this more complex configuration
until future times when more knowledge is acquired about the atmospheres of close-in super-Earths. The existence of such atmospheres
cannot currently be taken for granted even for the habitable zones \citep{heng}.

 \section{What comes first, solidus or peak heating?\label{sol.sec}}

 Investigating the dependence of the tidal heating rate on the average viscosity and temperature of Earth-like planets, \citet{beho} found that a
 positive feedback should lead to a runaway effect, i.e., to a situation where a higher mantle temperature causes even more heating. This can impact the habitability of close-in planets in the habitable zones of low-mass stars. For synchronously rotating planets, the possibility of runaway heating depends on the orbital eccentricity value and on the presence of plate tectonics, which strongly enhances the heat transfer in the mantle. The feedback is seen in our Figure \ref{dedt.fig} as the descending branch of the function on the right-hand side. Under synchronous rotation, the excitation frequency is (nearly) constant in time, but the effective viscosity and Maxwell time change with the rise of the temperature. An initially cold inviscid planet with a large Maxwell time begins to warm up if the heat transfer does not keep up with the energy input. Then, the viscosity decreases, and even more energy is dissipated. However, Figure \ref{dedt.fig} also shows that there is a self-regulation mechanism embedded in the Maxwell model. When the peak dissipation has passed, the steady warm-up and the shortening of the Maxwell time lead to a reduction of the damping rate. Using Figure \ref{kva.fig}, this can be visualized as the peak of the quality function shifting to the right as the viscosity drops and the value of the quality function is rapidly decreasing at a fixed frequency.

 A more radical and universal stop-mechanism for the runaway heating is provided by the phase transition of solid minerals when they begin to melt \citep{henh,sho}. Both the viscosity and the rigidity modulus precipitously drop, and the rate of dissipation is reduced by orders of magnitude. Even a limited fraction of partial melt is sufficient to bring about these drastic changes. This self-regulation mechanism is likely to have radically slowed down the initial expansion of the Moon's orbit by keeping the dissipation rate low despite the small orbital separation \citep{zah}. We do not expect even the closest exoplanets to be completely molten to the surface (magma ocean worlds)~---~at least, not by the tidal forces. However, since these processes can take place on potentially habitable planets orbiting M-type stars, the important question is whether the point of solidus is reached sooner than the peak dissipation, i.e., which of the self-regulation mechanisms halts the runaway internal heating.

 We analyzed three different rheological models proposed in the literature in application to the TR1 planets. These models assume a steeper, Arrhenius-type
 dependence of the mineral viscosity on temperature than the Frank-Kamenetskii approximation used by \citet[][;]{beho} \citep[the implications of this choice are discussed in, e.g.,][]{ste}. We find that the model proposed by \citet{dor}, which is limited to the range of temperatures $[1000,\;2500]\,^{\circ}\mbox{C}\,$, provides unrealistically low values of viscosity at higher temperatures that are expected in the deep layers of tidally heated exoplanets. The rheological model described by \citet{sta} gives more reasonable results when applied to the inner TR1 planets. For example, for the conditions expected at the bottom of the Earth's mantle ($T=4000$ K and $\,P=136$ GPa), the estimated viscosity is $\eta=1.8\times 10^{18}$
 Pa$\,\cdot\,$s. The corresponding Maxwell time is $264$~d, assuming an effective rigidity of $\mu=8\times 10^8$ Pa. We consider these values to be the benchmark solidus viscosity. In fact, the presolidus rigidity is also a function of temperature and pressure, but this dependence is much weaker than that for viscosity.

 A surprising result emerging from this model is the long Maxwell time for the presolidus deep mantle of the Earth and, consequently, a low rate of dissipation (see Section \ref{kva.sec}). However, this finding should be considered with due caution because of the large uncertainties associated with the physical parameters involved and the very strong dependences assumed in the model. For example, the estimated gradients of the Maxwell time at this $(T,P)$ phase point are $-1.1$~d~K$^{-1}$ and 14~d~GPa$^{-1}$. The resulting value may be lower by orders of magnitude if the actual solidus temperature is somewhat higher or if the actual pressure is somewhat lower. Second, the colder outer layers of the mantle can provide most of the dissipation if the pressure is significantly lower. Without detailed profiles of the temperature and pressure with depth, it is not possible to predict which layer is responsible for the bulk of the dissipation. Unfortunately, these conditions are not accurately known even for the Earth and are completely subject of speculative calculation for exoplanets. In particular, the solidus temperature versus pressure was experimentally estimated by \citet{hir}, but the model is confined to the range $[0, 10]$ GPa, which is hardly relevant to the tidally reacting deep layers of the mantle. A slightly more elaborate, but very similar, solidus model is found in \citet{tak}.

\begin{figure}[htbp]
  \centering
 %  \plottwo{Stamodel.eps}{Padmodel.eps}
  \includegraphics[scale=0.7]{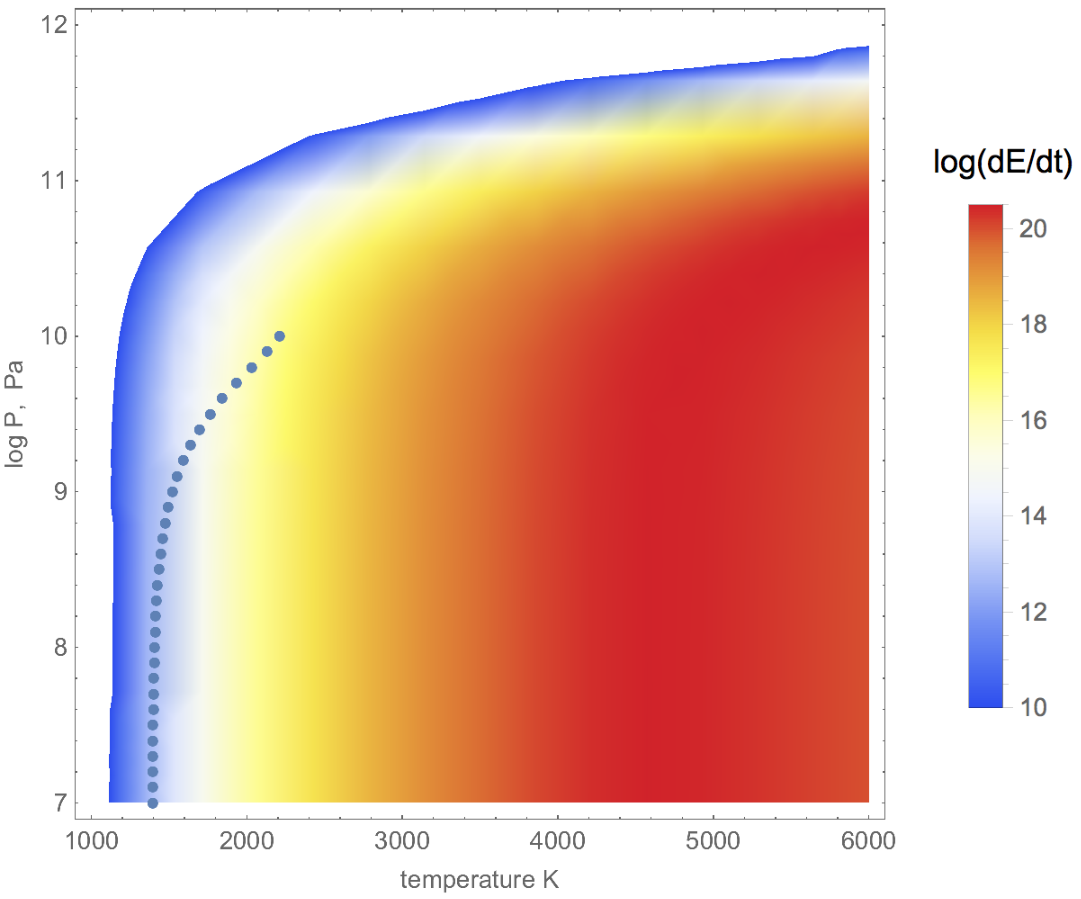} \includegraphics[scale=0.9]{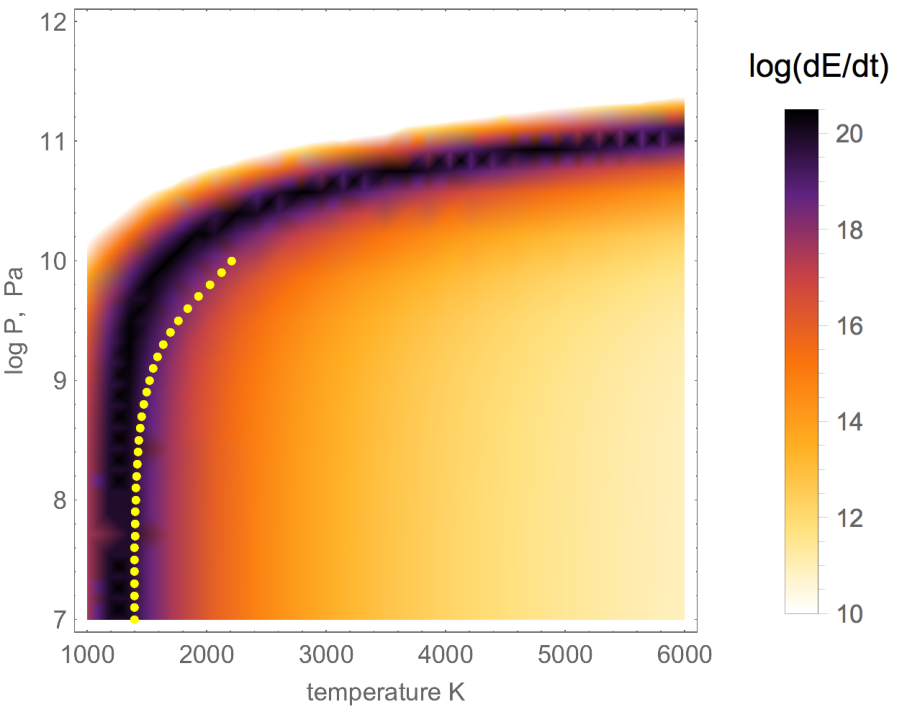}
 \caption{The rate of tidal dissipation as a function of temperature and pressure, for planet TR1-1. The incidence of solidus is
 shown with filled circles. Left: viscosity model of \citet{sta}. Right: Andrade-type model of \citet{pad}.
 \label{sta.fig}}
 \end{figure}

 Figure \ref{sta.fig} shows the color-coded rate of dissipation as a 2D function of temperature and pressure inside an Earth-like silicate mantle. For comparison, the expected solidus points from \citet{hir} are shown with filled circles. It appears that the model by \citet{sta} definitely predicts that solidus occurs before the peak tidal dissipation. In other words, the mantle begins to melt at much lower temperatures than those required to reach the peak dissipation rates within the Maxwell model. In this scenario, the entire mantle should be partially molten (perhaps, with a thin surface crust staying solid). This conclusion should be qualitatively valid for all the seven planets.

 A different picture may emerge from the model proposed by \citet{pad}. Those authors argued that the Maxwell model is not realistic for terrestrial-type solids with a grainy structure, while the Andrade mechanism of pinning-unpinning of defects might be a more appropriate model of the tidal friction mechanics. The frequency-dependent part of the Andrade friction can be formally characterized by an ``Andrade time'' parameter \citep{efr1}. The internal friction takes place when the grains start to shift relative to each other, which requires the external stress frequency to surpass a certain threshold value, see footnote \ref{3} and a reference therein. The threshold frequency exponentially grows with temperature and also depends on the characteristic grain size. The value of the threshold frequency remains a poorly constrained
parameter. In the model used by \citet{pad}, it is estimated directly from the ambient pressure and temperature,
without separately modeling the viscosity and rigidity modulus (S. Padovan, 2015, priv. comm.). Our calculations show that it is
possible to achieve the peak dissipation before the solidus for the inner TR1 planets (see Figure \ref{sta.fig}, right) if
a small grain size is assumed. For this calculation, we assumed a size of 1 $\mu$m (1 micron), which is probably too small.
Larger grains would also place the peak dissipation behind the solidus. We conclude from this analysis that the existing and
sufficiently elaborate models of exoplanet interiors suggest the termination of overheating by solidus, whence we expect the inner planets
of TR1 to be partially molten.

 \section{Can the inner planets be pseudosynchronous?}

 The combination of the models by \citet{sta} and \citet{hir} suggests that for the boundary condition $(T=2200$ K, $\;P=10$ GPa$)$, the presolidus Maxwell time is approximately 700 d. We do not know what happens in the deeper layers of the exoplanet's mantle. Despite this long time estimate, it is possible to simulate scenarios where the inner four planets are ``softened" by partial melt over the entire depths of their mantles. Indeed, if the heat transfer is inefficient due to the absence of plate tectonics, then~---~for a heat capacity of 1200 J kg$^{-1}$ K$^{-1}$ and a very conservative dissipation rate of $10^{16}$ W for a synchronously rotating TR1-1~---~the interiors should be warmed up by 50 K in just 1 Myr. The rise in temperature over 1 Gyr in this ``stagnant lid" scenario is $220\,000$, 54, 920, and 522 K for planets TR1-1 through 4; and is, at most, a few K for the others.
 Inevitably, the deep layers of planets 1, 3, and 4 will be consumed by the liquid core, and the outer layer will be partially molten. Such relatively inviscid, easily deformable planets are no longer governed by the solid-mineral rheology, but perhaps are more accurately described by the Maxwell model
 in the ``semiliquid" regime, characterized by the finite rigidity of the outer shell and a short Maxwell time \citep{ma15}. It can also be viewed as an approximation for a ``solidus-state" terrestrial exoplanet marginally heated by tidal forces.

 Cold and inviscid Earth-like planets cannot be captured in the so-called pseudosynchronous equilibrium, because the tidal quality ratio $\,k_2/Q\,$ declines in magnitude with increasing perturbation frequency in the vicinity of a spin-orbit resonance \citep{mae}. On the other hand, within the Constant Time Lag model (where the quality ratio $\,k_2/Q\,$ is linear in the perturbation frequency), pseudosynchronism is inevitable for any nonzero eccentricity
 \citep{mig, hut81, mur, we, rod}. A semimolten state may be implemented between these two extremes, with the Maxwell time being short enough for the peak dissipation frequency to be higher than the perturbation frequency. As was shown by \citet{ma15}, the main condition for pseudosynchronism is that the magnitude of the secular acceleration caused by the residual tidal dissipation should be greater than the amplitude of free libration (taking the maximum in the vicinity of the 1:1 spin-orbit resonance). Practically, this means that the eccentricity should be large enough and the triaxiality small enough for the planet to retain a slightly faster equilibrium rotation rate than the synchronous value. The two conditions can be presented as
  \begin{eqnarray}
  \tau_M &\la &\frac{1}{n(1+{\cal A}_2)}\,\;,
  \label{condition1}\\
  \nonumber\\
  e &> & 0.43\;\sigma^\frac{1}{4}\,\;,
  \label{condition2}
  \end{eqnarray}
 where $\,\sigma=(B-A)/C\,$ is the triaxiality parameter computed from the principal moments of inertia. The dimensionless rigidity $\,{\cal A}_2\,$ is given by
 \ba
 {\cal{A}}_2=\,\frac{\textstyle 57}{\textstyle 8\,\pi}~\frac{\textstyle \mu}{\textstyle
 G\,\rho^2\,R^2}\,\;
 \label{A2}
 \ea
 where $\,G\,$ is the Newton gravitational constant,
 while $\,R\,$, $\,\rho\,$, and $\,\mu\,$ are the radius, the mean density, and the mean shear rigidity of the body, correspondingly.

 Condition (\ref{condition1}) results in the upper limits of $\,(0.05,0.15,0.12,0.16,0.22,0.69,0.82)$ d for planets 1 through 7, respectively. Although much shorter than our estimated solidus Maxwell time at 10 GPa (700 d, cf. Section \ref{sol.sec}), such values cannot be precluded for partially melted mantles. Maxwell times shorter than 1 d can be expected for semiliquid ice satellites of the Solar System, e.g., Io or Enceladus.

 Condition (\ref{condition2}) is more definitive. The moments of inertia are not known, but we can compute the upper limits from the estimated eccentricities and compare with the Earth's value. We find that the only planets that pass this criterion are TR1-1 and TR1-4, although the upper limit for TR1-3 ($1\times 10^{-5}$) is only a half of the Earth's value. The same three planets (1, 3, and 4) have the highest probability of capture into the 3:2 resonance (which would preclude the pseudosynchronous state). However, according to Figure \ref{pcapt.fig}, this can only happen at much longer Maxwell times.

 Thus, planets TR1-2, 5, 6, and 7 are likely to reside in synchronous rotation states, remaining cold, inviscid, and undifferentiated. If the planets 1, 3, 4 were cold at the time of initial spin-down, they could be captured into a higher-than-synchronous resonance with the ensuing rapid internal heating and partial meltdown.

 \section{Escape from the 3:2 resonance}

 What happens with a planet trapped in a 3:2 spin-orbit resonance, heated by tidal friction? \citet{noel} numerically investigated spin-orbit evolution scenarios for Mercury under variable eccentricity conditions and concluded that a planet practically never leaves a resonance, once captured. This observation was made for a ``cold" Maxwell model with a fixed average viscosity. A close-in exoplanet locked in the 3:2 or higher resonance receives almost two orders
 of magnitude more of tidal heating (Figure \ref{dedt.fig}). The rise of internal temperature in combination with a stable pressure brings about a drop in viscosity of several orders of magnitude (Section \ref{sol.sec}). As the dominant quality peak corresponding to the 1:1 resonance shifts toward the 3:2 resonance and the 3:2 kink spreads out, the necessary condition for the 3:2 circulation is no longer fulfilled. We then expect the planet to leave the 3:2 spin-orbit resonance by itself, without any additional perturbation, and start its descent to the 1:1 resonance.

\begin{figure}[htbp]
    \begin{center}
    \includegraphics[scale=1.7]{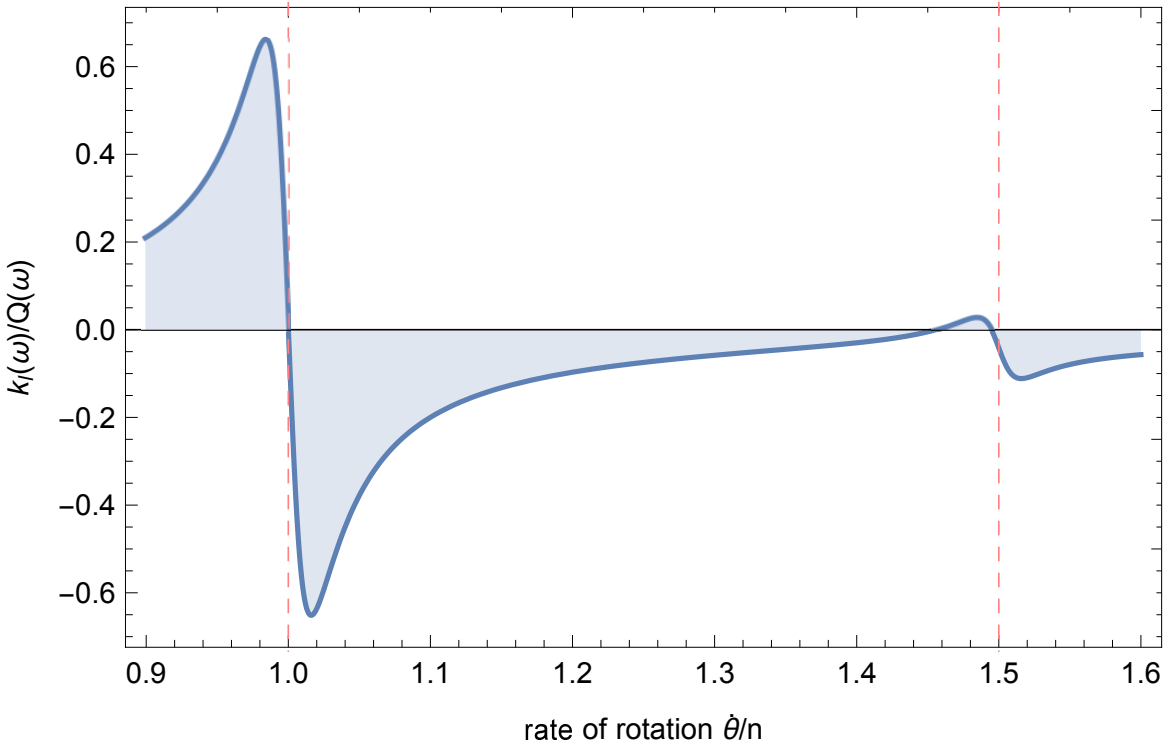}
    \caption{
    \label{kinks.fig}
Overall tidal quality function for the planet TR1-1 (b) presented as a function of the spin rate $\,\dot{\theta\,}$ treated as a slowly changing parameter. The vertical dashed lines mark the spin-orbit resonances.
    The principal, large kink corresponds to the semidiurnal term with $\,\{lmpq\}\,=\,\{2200\}\,$, while the small kink corresponds to the term with $\,\{lmpq\}\,=\,\{2201\}\,$. At lower temperatures, the small kink is crossing the horizontal axis and is therefore acting as a trap, ensuring that the planet gets caught in the 3:2 spin-orbit resonance. As the temperature increases, both kinks spread out. As soon as the smaller kink ``sinks'' below the horizontal axis, it can no longer act as a trap, and the body begins to decelerate 
toward synchronism (or pseudosynchronism, if it is semimolten).}
    \end{center}
    \end{figure}

 We now recall that a tidal Fourier mode $\,\omega_{lmpq}\,$ excited in a rotating body is a linear combination of the mean motion $\,n\,$ and the angular velocity $\,\stackrel{\bf\centerdot}{\theta\,}$ of the body \citep{ema}:
 \ba
 \omega_{lmpq}\,\approx\,(l-2p+q)\,n\,-\,m\,\dot{\theta}
 \,\;,
 \label{mode}
 \ea
 while the tidal torque acting on the body is an infinite sum, over $\,lmpq\,$, of terms, each of which is proportional to the quality ratio
 \ba
 \frac{k_l(\omega_{lmpq})}{Q_l(\omega_{lmpq})}\,\;.
 \label{quality}
 \ea
This ratio, as a function of $\,\omega_{lmpq}\,$, has the shape of a kink, such as that seen in Figure \ref{kinks.fig} around the
$\dot\theta=n$ resonance.
 The tidal quality ratio $\,{k_l(\omega_{lmpq})}/{Q_l(\omega_{lmpq})}\,$ has this shape for any rheology. The shape is natural: on the one hand, the torque component must go smoothly through zero on crossing a spin-orbit resonance; on the other hand, the material's reaction has some delay and cannot respond instantaneously to forcing at the frequencies much higher than the Maxwell time~---~hence the fall-off at high frequencies \citep{efr1, efr2}.

Taking the weighted sum over all tidal modes, treating $\,n\,$ as a slowly changing parameter, and expressing all the terms of the 
quality function as functions of $\,\stackrel{\bf\centerdot}{\theta\,}$, we obtain the overall quality function of frequency, which looks like a superposition 
of kinks of different amplitudes in Figure \ref{kinks.fig}.
 In this figure, a smaller kink (corresponding to the term $\,\{lmpq\}\,=\,2201\,$) is residing on the slope of the principal, semidiurnal term. 
The capture into the 3:2 spin state takes place when the small torque crosses the horizontal axis. Now consider tidal despinning of an oblate symmetrical planet with no triaxiality. If we start close to the 3:2 spin-orbit state (but with a slightly faster initial spin), we are guaranteed to get trapped into this state, provided the smaller kink is crossing the horizontal axis. In this case, the small kink is acting as a trap, ensuring that the oblate rotator gets captured in the 3:2 spin-orbit resonance. As the temperature increases, the Maxwell time
becomes shorter, and both kinks spread out. At some point, the smaller kink ``sinks'' below the horizontal axis. The secular tidal torque
becomes negative everywhere in the vicinity of the 3:2 resonance, and the body begins despinning toward synchronism (or pseudosynchronism, if semimolten).

\begin{figure}[htbp]
  \centering
  \plotone{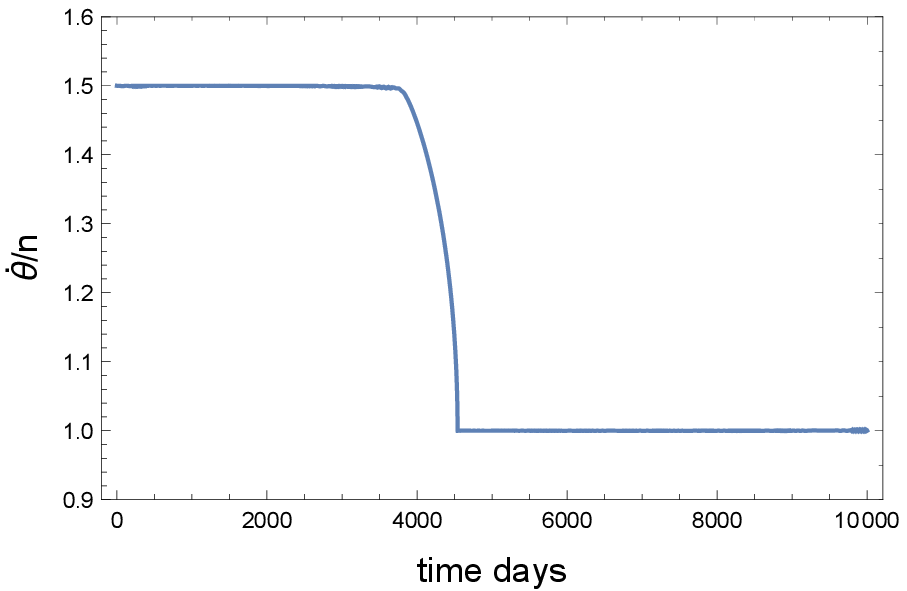}
\caption{Simulated rate of rotation of planet TR1-1 initially captured in the 3:2 spin-orbit
resonance, with the Maxwell time exponentially dropping from 100 d to 0.1 d by the end of the time span. \label{jump.fig}}
\end{figure}

 To confirm this scenario, we performed numerical simulations of planet TR1-1 captured in the 3:2 spin-orbit resonance, subject to the regular triaxial torque and the tidal torque from the host star. The differential equation of motion for the spin of this planet was integrated for 10000 days with a maximum step of $3\times10^{-4}$~d. The shortening of the Maxwell time was artificially accelerated, modeled as an exponential decline starting at $\tau_M=100$~d at
 $t=0$ and ending at 0.1~d at $t=10000$~d. All other essential parameters were kept fixed ($M_2=0.88\;M_E$, $R=1.087\;R_E$, $e=0.03093$, $P=1.5108708$~d, $\sigma=1.9\times10^{-5}$). Figure \ref{jump.fig} shows the resulting trajectory for the relative rotation velocity, $\dot\theta/n$. The planet spontaneously jumps out of the 3:2 resonance and very quickly descends into the 1:1 resonance. Note that the rate of spin-down was not artificially accelerated here,
 being defined by the magnitude of tidal dissipation.

 The planet breaks out of the 3:2 lock in the middle of this simulation, when the Maxwell time is approximately equal to 8~d. This is still a very ``cold" model, with the tidal quality ratio $\,k_2/Q\,$ peaking at a low frequency (cf. Figure \ref{kva.fig}). As we noted before, when both the eccentricity and the triaxiality are small, a stable spin in a higher-than-synchronous resonance is possible only when the secular tidal torque takes positive values in the close vicinity of the resonant frequency. The sign of the tidal torque near the 3:2 resonance is defined by the balance of the two leading inputs, those corresponding to the Fourier tidal modes $\,\omega_{2200}\,$ and $\,\omega_{2201}\,$.

Thus, the escape from the resonance happens when the peak acceleration related to the $\omega_{2201}$ mode becomes smaller in magnitude than the
secular acceleration associated with the synchronous $\omega_{2200}$ mode because of the decreasing Maxwell time. For a small $\,e\,$, this happens,
to a good approximation, at
\eb
K(n)\;=\;\frac{3}{2}\;\frac{{\cal A}_2\;n\;\tau_M}{(n \tau_M)^2\;(1+{\cal A}_2)^2\,+\;1}=\frac{147\;{\cal A}_2\;e^2}{16\;(1\,+\,{\cal A}_2)}\,\;.
\ee
The larger root of the emerging quadratic equation yields
\eb
^{(escape)}\tau_{M}\,=\;\frac{4+\sqrt{16-2401\,e^4}}{49\;n\;e^2\,(1+{\cal A}_2)}.
\ee
This formula is only valid for $e <0.2857$. At greater eccentricities, a planet can not escape the 3:2 resonance by itself within
the framework of the Maxwell model. The calculated escape value for the TR1-1 planet is $8.25$ d.

 \section{Tidal dissipation in the host star}

 Most of the exoplanet hosts, being Gyr-old stars, rotate slower than the close-in planets \citep{mat}. The tidal bulge raised on the surface of the star by a planet lags behind the direction to the planet. The situation is opposite to that in the Earth--Moon system, where the tidal bulge running across the surface of the Earth leads the direction to the Moon. The lagging tidal bulge inevitably accelerates the star's spin and reduces the system's eccentricity and orbital distance.

 Photometric light curves obtained with different instruments lead to discrepant estimates for the TRAPPIST-1 star rotation period, such as 3.30~d \citep{vid,lug}, 1.40~d \citep{gil16}, and 0.819~d \citep{roe}. The shorter periods may be caused by a group of major photometric structures distributed in longitude. The longer period, on the other hand, seems to be in disagreement with the measurement of the projected velocity \citep{rei}, which suggests a value not too different from 1~d. At any rate, only the two inner planets TR1-1 and 2 may have a higher mean motion than the host's rotation rate. The other five planets move slowly, and their orbits may expand and become more eccentric. However, the action of these planets on the star is vanishingly small compared to the inner planets. The direction of the stellar spin evolution is mostly defined by the tidal action of the innermost planet. In the long term, this evolution will reflect on the orbits of the outer planets. Thus, the innermost planets change the orbits of the outer planets through the medium of the tidal response on the host star, and over a very long term, this action may be more significant than the direct gravitational interaction between the planets.

 Using the potential expansion by \citet{kau},
 %  further developed by \citet{rem},
 we derived the orbital evolution equations for the special case discussed here: ~(1) a small eccentricity, $\,e\to 0\,$; ~(2) a negligible obliquity of the orbit on the stellar equator, $\,i_1= 0\,$; ~(3) the star's rotation is not synchronized, $\,\dot{\theta}_1 \ne n\,$. $\,$To the most significant power of $\,e\,$, the resulting differential equations are
 \begin{eqnarray}
 \frac{d a}{d t} & \simeq& -3 n\,a\,\left(\frac{M_2}{M_1}\right)\left(\frac{R_1}{a}\right)^5\,
 K_1(2n-2\dot{\theta}_1)
 \label{das.eq}\\
 \frac{d e}{d t} & \simeq& - n\,e\,\left(\frac{M_2}{M_1}\right)\left(\frac{R_1}{a}\right)^5\, \times \nonumber\\
 && \left(-\frac{3}{16}K_1(n-2\dot{\theta}_1)-\frac{3}{4}K_1(2n-2\dot{\theta}_1)+\frac{147}{16}K_1(3n-2\dot{\theta}_1)+\frac{9}{8}K_1(n)\right)\,\;.
 \label{des.eq}
 \end{eqnarray}
 Here the subscript ``1'' refers to the star and ``2'' to the planet. Accordingly, $M_1$ and $M_2$ are the stellar and planetary masses, while $\theta_1$ and $\dot{\theta}_1$  denote the polar rotation angle and the polar rotation rate of the star. The notation
 \ba
 K_1(\omega_{2mpq})\;\equiv\;\left[\,\frac{k_2(\omega_{2mpq})}{Q(\omega_{2mpq})}\,\right]^{(star)}
 \label{quality_star}
 \ea
 stands for the quality ratio $\,k_2/Q\,$ of the star, which is a function of the tidal Fourier mode $\,\omega_{2mpq}\,$ excited in the star; see expression (\ref{mode}).  In our approximation, we need to take into consideration the inputs with $\,\omega_{201,-1}\,=\,-\,n\,$, $\;\omega_{2011}\,=\,n\,$, $\;\omega_{220,-1}\,=\,n\,-\,2\,\dot{\theta}\,$,
 $\;\omega_{2200}\,=\,2\,n\,-\,2\,\dot{\theta}\,$, $\;\omega_{2201}\,=\,3\,n\,-\,2\,\dot{\theta}\,$.

Expressions (\ref{das.eq}) and (\ref{des.eq}) are general and valid for an arbitrary rheological model. The quality function (\ref{quality_star}) is an odd function of the Fourier mode,
  $$
  \mbox{Sign}[\,K_1(\omega_{2mpq})\,]\,=\,\mbox{Sign}(\omega_{2mpq})\,\;.
  $$
  Also be mindful that, irrespective of the tidal model, $\,K_1(0)=0\,$. \,This condition warrants the crossing of spin-orbit resonances without experiencing an abrupt change of the tidal torque value.

 We still do not know the sign of the principal (semidiurnal) tidal mode $2n-2\dot{\theta}$. In other words, we do not know whether the star rotates faster or slower than the orbital motion of TR1-1. Depending on that sign, the inner planet may spiral in or out, per equation (\ref{das.eq}). Since the other planets are locked in a complex chain of MMRs, they are expected to follow suit. The coefficients in front of the quality functions for planets TR1-1 and 2 are 0.11 and
 0.03~m~d$^{-1}$. Tidal dissipation is most efficient when $\,K_1(\omega_{2mpq})\,$ attains its highest possible values, which are of the order of unity. In this case, the shortest characteristic times of the distance evolution $\,a/|\dot{a}|\,$ are of the order of $\,10^7$--$\,10^8$ yr. Thus, in principle, the tides on the host star might have impacted the orbital configuration had the tidal quality ratio assumed sufficiently high values. It is, however, commonly believed that the values of $\,k_2/Q\,$ for stars are orders of magnitude smaller than unity. If this assumption is true, then the role of the stellar tides in the orbital dynamics is small, if any.

 The situation with the sign of $\,de/dt\,$ is even less certain, because in equation (\ref{des.eq}) four tidal Fourier modes provide contributions of the order of $\,e^1\,$. The overall coefficients in front of the linear combination of the quality terms amount to $1.4\times 10^{-12}$ and
 $1.2\times 10^{-14}$~d$^{-1}$ for planets TR1-1 and 2, respectively. The corresponding characteristic times $\,e/|\dot{e}|\,$, to the precision of the unknown quality factors, are also of the order of $\,10^8$ yr. It is almost certain that the sign of this derivative is negative, i.e., that the tidal action tends to circularize the orbit. We conclude again that the tidal effect on the host star is unlikely to impact much the orbital configuration, unless the value of the tidal quality function of the stellar material is high~---~which is against our expectations.

 Finally, the angular acceleration of the star due to the tidal dissipation in the star is, to a good approximation in this case, given by
 \eb
 \ddot{\theta}_1  \simeq \frac{3}{2} \frac{n^2}{\xi}\,\left(\frac{M_2}{M_1}\right)^2\left(\frac{R_1}{a}\right)^3\,
 K_1(2n-2\dot{\theta}_1)\,\;,
 \label{}
 \ee
 with $\xi$ denoting the moment of inertia coefficient (equal to 0.4 for a uniform sphere). The corresponding characteristic times $n/|{\ddot{\theta}}_1|$ of the spin-down (or spin-up for $n>\dot{\theta}_1$) are considerably longer, being of the order of 1 Gyr at a minimum. This simple calculation demonstrates that the planets are too small to appreciably change the rate of rotation of the TRAPPIST-1 star. The observed quick spin of the star therefore represents a puzzle, unless the star is quite young.

 \section{Orbital evolution due to the tides in a planet\label{orb.sec}}

 Following the same procedure, we start with the Kaula expansion for the disturbing potential, inserted in Lagrange orbital equations. This gives us the following approximations for the rates of $\,a\,$ and $\,e\,$ caused by the tidal dissipation in a planet:
 \begin{eqnarray}
 \frac{d a}{d t} & \simeq& -\frac{447}{8} n\,a\,e^2\,\left(\frac{M_1}{M_2}\right)\left(\frac{R_2}{a}\right)^5\,
 K_2(n) \label{dap.eq}\\
 \frac{d e}{d t} & \simeq& - \frac{21}{2} n\,e\,\left(\frac{M_1}{M_2}\right)\left(\frac{R_2}{a}\right)^5\,\;
 K_2(n)\label{dep.eq}
 \end{eqnarray}
 As before, $M_1$ and $M_2$ are the stellar and planetary masses. The notations $\theta$ and $\dot{\theta}$ stand for the polar rotation angle and the polar rotation rate of the planet, while $\,K_2(n) = k_2(n)/Q(n)\,$ is the tidal quality ratio (the tidal quality function) of the planet, taken at the Fourier tidal mode equal to $1\,n$. These expressions can be used for the
 specific configuration when \,(1) the orbital eccentricity is small; (2) the orbital obliquity $\,i\,$ on the planetary equator is negligible; (3) the planet
 is synchronized ($\dot{\theta} = n$). Note that the rate $da/dt$ is now proportional to $e^2$, because in the 1:1 spin-orbit resonance all the terms of the zeroth order in $\,e\,$ become zero.

 The greatest uncertainty in applying this formula to the TR1 system comes from the unknown value of the tidal quality functions of the star and of the planets. To the precision of these unknown values, the characteristic times of the orbital decay, $\,a/|\dot{a}|\,$, amount to 0.3 Myr, 2 Gyr, 0.1 Gyr, and 0.2 Gyr for planets
 TR1-1 through 4, respectively. It may seem puzzling that the tides on the innermost planet may be much more efficient at shrinking the orbit than the tides on the star, albeit the explicit proportionality is to the fifth power of the perturbed body's radius. The reason for this is the mass ratio, which is inverse and outweighs the impact of the smaller bulge. Qualitatively, this happens because a smaller perturber's mass generates a smaller bulge and tidal torque, but the angular momentum gets distributed to the total mass of the system anyway. It should be noted here that in all the preceding derivations, we assumed $M_1\gg M_2$. Unlike the case with stellar tides, a planet's quality ratio $\,k_2/Q\,$ can be quite large in this case, up to the absolute maximum of $3/4$ at the peak dissipation
 frequency. Therefore, depending on the efficiency of the tidal friction, the tides in a planet can be a significant orbital evolution factor, especially for the inner planets. Assuming a Maxwell time of 1 d for all of the planets, we estimate the characteristic times of orbital decay, presented in the second column of Table \ref{cha.tab}. It is possible that the angular momentum loss for the innermost planet is redistributed in the entire system by means of the MMR chains \citep{pap}, leading to rather complex dynamical scenarios, which are outside of the scope of this project.

 The situation is different for the rate of circularization caused by the tides on the synchronized TR1 planets. Even though the rate of dissipation is at a global minimum in this state \citep[except for the possible pseudosynchronism of semiliquid bodies; see][]{lev,wis}, the derivative $de/dt$ is still $O(e)$, and takes rather large values for the inner planets. Table \ref{cha.tab}, third column, gives the characteristic times of circularization, computed for
 $\tau_M=1$~d. The innermost planet, left to itself, would have completely circularized in a matter of $30\,000$ years. We, however, know that this planet has a small but measurable eccentricity. Almost certainly, the finite eccentricity of the TR1 planets is excited by the chaotic gravitational interactions boosted by the MMR chains. We have seen in our numerical simulations that the system breaks up when a planet abruptly increases the eccentricity and plunges into the star. Hence, the tidal damping of eccentricity may be pivotal in the long-term survival of the system, providing especially strong stabilization to the
 most vulnerable inner planets. Still, the lifetime trend for the system is to shrink. The relatively short decay time for the TR1-1 planet is probably not implemented in reality, because it is not possible for it to leave the 5:3 MMR with TR1-2, and other, weaker resonances.

 \begin{deluxetable}{lrr}
 \tablecaption{Characteristic times of the orbital decay ($a/|\dot{a}|$) and circularization ($e/|\dot{e}|$) in Myr, assuming $\tau_M=1$ d.\label{cha.tab}}
 \tablewidth{0pt}
 \tablehead{
 \multicolumn{1}{c}{Planet}  & \multicolumn{1}{c}{$a/|\dot{a}|$} & \multicolumn{1}{c}{$e/|\dot{e}|$} \\
 \multicolumn{1}{c}{} & \multicolumn{1}{c}{Myr} &
 \multicolumn{1}{c}{Myr}\\}
 \startdata
 TR1-1 & $5.6$ & $0.03$\\[1ex]
 TR1-2 & $>10000$ & $0.17$\\[1ex]
 TR1-3 & $710$ & $2.2$\\[1ex]
 TR1-4 & $960$ & $5.2$\\[1ex]
 TR1-5 & $>10000$ & $15$\\[1ex]
 TR1-6 & $>10000$ & $43$\\[1ex]
 TR1-7 & $>10000$ & $730$\\[1ex]
 \enddata
 \end{deluxetable}

 \section{Direction for further research: eccentricities coupling and its possible role in stabilizing the system}

 The formulae for $\,da/dt\,$ and $\,de/dt\,$ presented in the above two sections are those of a two-body problem. In a multibody setting, however, various mutual interactions come into play. In the absence of mean-motion resonances, the long-term evolution of a conservative multiplanetary system is described, in the first order, by a system of the Laplace--Lagrange linear secular equations \citep{laskar1990}. Within this approximation, inclinations and eccentricities are decoupled. Orbital elements become coupled when dissipative phenomena (tides) are taken into account. In this situation, one should consider the eigenmodes of eccentricity, as was done in \citet{greenberg} and \citet{laskar}. Things are elevated to a further level of complexity by MMRs. While on physical grounds we expect that under MMR the couplings are strengthened, mathematical derivation of the Laplace--Lagrange system becomes more difficult. Its integration will become problematic because, as we have found, the behavior of eccentricities demonstrates chaos.

 While these effects are beyond the scope of our current paper, we wish to point them out, because they may provide a key to how the TRAPPIST-1 system ended up on its island of stable solutions.

 \section{Discussion and conclusions}

 We have carried out the numerical integration of the long-term evolution of the TRAPPIST-1 system, starting with a previously suggested stable configuration.
 Having confirmed the long-term stability of this configuration, we also have found that the motion of planets is chaotic. The eccentricity values vary within certain ranges in an apparently random way. Using the robust mean values of eccentricity, the expected spin-orbit states have been computed in the probabilistic sense. The rates of tidal dissipation and tidal evolution of orbits have been estimated, assuming an Earth-like rheology for the planets.

 The TRAPPIST-1 system may not be an oddball in the world of planetary systems, but a rather common and natural outcome of star formation and $N$-body dynamics, where tidal interactions undoubtedly play a big role. The orbital motion of the tightly packed planets is distinctly chaotic, with a ``short-fused'' Lyapunov time~---~and this picture may be normal for exoplanet systems. The complex multiplet of MMRs, where each of the seven planets has a commensurable orbital motion with each of the other six planets, suggests a highly organized structure unseen in the Solar System. The challenge is to explain how this degree of organization and complexity can naturally emerge from the primary stochastic material, following the principles of classical mechanics and thwarting the flow of entropy. The MMR chains have a regularization effect on the orbital dynamics, damping the chaotically emerging perturbations and keeping eccentricity at low levels. This prevents the planets from coming at close
range and directly interacting with each other. Are there other regularization agents? Could the secular tidal interaction have made the long-term orbital evolution more orderly and constrained?

 Working within the Maxwell rheological model for the planets, we find that planets TR1-1, 3, and 4, with initial fast rotation in the prograde sense and
 cold inviscid mantles, are certain to be originally captured into the 3:2 or even higher spin-orbit resonance. At the current values of their orbital eccentricities, planets 6 and 7 could have been captured too, if their average viscosity was high enough. In higher-than-synchronous resonances, the rotational dynamics with tides is quite complicated and nonlinear, with higher levels of forced libration involved \citep{fe} and even with the possibility of additional secondary attractors in the phase space \citep{bar}. The rate of the kinetic-energy dissipation is $O(e^0)$ for all higher-than-synchronous resonances, which brings about high rates of internal heating. For TR1-1, for example, the heating rate in the 3:2 spin state is at least an order of magnitude higher than in the  synchronized state and can reach $\,10^{22.5}$ W when the Maxwell time value corresponds to the peak of the quality function (see Figure \ref{kva.fig}).
 The emerging equilibrium surface flux amounts to a few MW m$^{-2}$. For comparison, it is equivalent to a small power-generating gas turbine on each square meter of the planet surface. Enceladus, one of the most impressive tidal power plants in the Solar System, generates $\sim 10^{10}$ W, which corresponds to the mean surface flux of $\sim 10^{-2}$ W m$^{-2}$ \citep{kamata, efr4}. The power output on more distant planets of TRAPPIST-1 is less extreme, but quite significant, too. The internal temperature inevitably rises, as the tidal heat cannot be efficiently transferred to the surface and radiated out.

As the viscosity $\,\eta\,$ of silicate minerals is a strong function of temperature, the planet's interiors have lower values of $\,\eta\,$~---~which further increases the dissipation rate. Runaway heating appears inevitable \citep{beho}, turning the initially rocky planets into balls of liquid magma. We posit, however, that there are heating self-regulation mechanisms terminating the runaway process much sooner. First, at the low orbital frequencies observed for TR1 system, and with presumably small degrees of triaxial deformation, the condition of capture into the 3:2 spin state vanishes abruptly, and the planet jumps out of the resonance. It is bound to spin down very quickly, getting trapped in the permanent 1:1 spin-orbit resonance. The planets can be in the semiliquid state during this transition, so the pseudosynchronism equilibrium is possible in principle; however, only the inner planet passes the eccentricity criterion of pseudosynchronism. The rate of dissipation drops by an order of magnitude (cf. Figure \ref{dedt.fig}). This is still quite high for the inner four planets, and they continue to warm up. The ultimate stop to this process is set by the instance of solidus, when part of the material starts to melt. Both the viscosity and the rigidity modulus take a dive by many orders of magnitude, resulting in ultra-short Maxwell times and, hence, a much lower dissipation rate. According to the existing models, the solidus point should be $\sim 1400$ K at the ambient pressure of 10 Gpa (though we do not have good knowledge on the deeper parts of the mantle), which implies that the melt starts to grow well before the peak dissipation of the Maxwell model is achieved (peak in the sense of Figure \ref{kva.fig}). Due to the strong negative feedback between the amount of partial melt and tidal dissipation, an equilibrium should be achieved, setting the temperature profile and the power output in balance with the efficiency of the surface energy loss. If a planet in this equilibrium has a
 thick Venus-like atmosphere and is devoid of plate tectonics, a larger amount of melt may be expected.

 While some of the planets ($\,b\,$, $\,d\,$, and $\,e\,$) were captured in the 3:2 or higher spin-orbit resonances during the initial spin-down,
the end product of this scenario is a TR1 system where all the planets rotate synchronously with their orbital motion, minimizing the tidal heating.
 A possible exception may be the closest planet $\,b$, which may, depending on its rheology, be trapped in pseudosynchronism.
 The inner four planets are likely to have massive molten cores and hot mantles with a significant degree of melt. These conditions help them to generate as much tidal heat as can be transferred to the surface and irradiated into space. The limited tidal dissipation in the planets means a limited impact on the orbital dynamics and stability of the system.
 Owing to the tidal dissipation inside the star, the orbits are expected to keep shrinking even in the synchronous state, though at a moderate or low rate (unless the tidal quality of the star is much higher than what is commonly assumed). The eccentricity of the planets, on the other hand, can go down quickly enough on the lifetime scale. We conclude that, for long-term stability to be maintained, the planet-planet gravitational interaction should keep the eccentricities' values  within finite ranges. We also conclude that the tides on the inner planets
can provide an additional regularization mechanism adding to the stability of the configuration.

\end{document}